# Intensity Correlated Spiking Emission Due to Cooperative Effects in Alkali Vapors

**Alexander M. Akulshin[1,2], Nafia Rahaman[1], F. Pedreros Bustos[2], Sergey A. Suslov[3], Russell J. McLean[1], and Dmitry Budker[2,4]**

[1] *Optical Sciences Centre, Swinburne University of Technology, PO Box 218, Hawthorn 3122, Australia*
[2] *Helmholtz-Institut, GSI Helmholtzzentrum für Schwerionenforschung, Johannes Gutenberg University, 55128 Mainz, Germany*
[3] *Department of Mathematics, Swinburne University of Technology, PO Box 218, Hawthorn 3122, Australia*
[4] *Department of Physics, University of California, Berkeley, CA 94720-7300, USA*
*Author e-mail address: (aakoulchine@swin.edu.au)*

**Abstract:** Spiking behavior and a high degree of intensity correlation of frequency up- and down-converted directional radiation from population-inverted alkali vapors excited with a continuous-wave laser pumping are attributed to cooperative effects. © 2021 The Authors

## 1. Introduction

Frequency conversion of resonant laser light in alkali vapors into directional radiation in THz to UV spectral regions continues to attract a lot of attention, motivated by possible applications ranging from quantum telecommunication and medical imaging to remote magnetometry [1,2]. To optimize the parameters of new optical fields to meet the requirements of a variety of applications, a detailed study of the most important processes involved is needed.

Recently, studying the temporal dynamics of collimated blue light (CBL) at 420 nm generated by FWM in Rb vapor, we found that despite cw excitation CBL consists of short irregular spikes [3]. The duration of individual spikes is less than the natural lifetime of both the excited levels. This spiking behavior was attributed to temporal features of the directional infrared emission generated on the population-inverted $5D_{5/2}$-$6P_{3/2}$ transition. It is well-known that population-inverted atoms can relax to a lower state much faster radiating a short burst of directional radiation [4,5]. Thus, the temporal properties of the mid-infrared radiation might be determined by cooperative effects and the observed spiking emission at 420 nm is the result of transferring the temporal features through the FWM process. To confirm this idea, we performed an experiment with Na vapors, since new fields generated on cascade $4D_{5/2}$ - $4P_{3/2}$ - $4S_{1/2}$ transitions (Fig.1a), are in the spectral regions where fast detectors are readily available.

## 2. Experimental results

When a sufficient number of population-inverted atoms is contained in the elongated region defined by the overlap of the pump beams the directional radiation is generated on two infrared transitions [6]. Despite the cw excitation, the temporal profiles of the 2338 and 2207 nm fields consist of chaotic, partially overlapping spikes. The spike appearance time, as well as their shape, duration and amplitude, vary considerably from sample to sample.

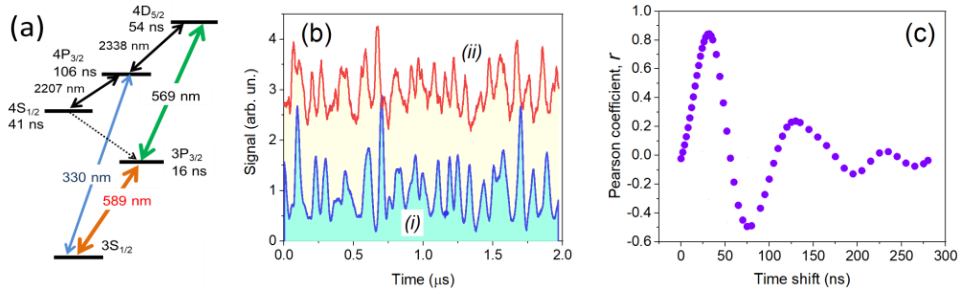

Fig. 1. (a) Energy levels of Na atom involved into new-field generation. (b) Temporal profiles of directional emission at *(i)* 330 nm and *(ii)* 2338 nm, respectively. The fixed-frequency pump lasers are tuned to the $3S_{1/2}(F=2)$-$3P_{3/2}(F'=3)$ and $3P_{3/2}(F'=3)$-$4D_{5/2}$ transitions. Curve *(ii)* is shifted vertically by two units for better visibility. (c) Pearson correlation coefficient *r* as a function of the time shift between profiles *(i)* and *(ii)*.

Also, the subsequent FWM of the forward-directed radiation at 2338 nm and the applied laser pump produces directional radiation at 330 nm. We note that intensity profiles of the infrared and UV radiation are remarkably similar, as shown in Fig. 1b. The Pearson coefficient *r*, which characterizes linear correlation between the variables, was calculated for these profiles a function of the time shift between them (Fig. 1c). This dependence shows not only a high degree of correlation ($r_{MAX} = 0.85$ at $\tau = 32$ ns), but also a remarkable form of damped oscillations. These oscillations reflect some periodicity of the spike occurrence.

Correlation between the fields at 2207 and 330 nm is weaker ($r_{MAX} = 0.61$) as these fields are not linked via the FWM process; however, its existence proves intensity correlation between the cascade-generated infrared fields.

Thus, these observations support the idea that spiking behavior of the CBL observed in Rb vapors is due to the temporal features of the directional emission at 5.23 µm, as this emission is generated on the population-inverted transition and possesses the most characteristic properties of the collective emission.

A further study of the intensity correlation was carried out with Rb vapors.

We find that intensity correlation of CBL fields generated simultaneously in the same cell in two spatially separated regions (Fig. 2a,b) depends on the geometry of the applied laser pumps. When a pair of parallel bi-chromatic pump beams are more than 4 mm apart, while their cross section is approximately 0.4 mm, intensity noise of two CBL fields is completely uncorrelated. At a smaller separation, < 3 mm, the intensity of each field decreases slightly if the neighboring beam is present, while the calculated Pearson coefficient for their intensity noise averaged over a set of profiles, shows weak anticorrelation, $r = -0.06 \pm 0.02$. However, if the applied pump beams are crossed at small angle (~50 mrad), the intensities of two CBL beams show a significant correlation, $r = 0.41 \pm 0.08$.

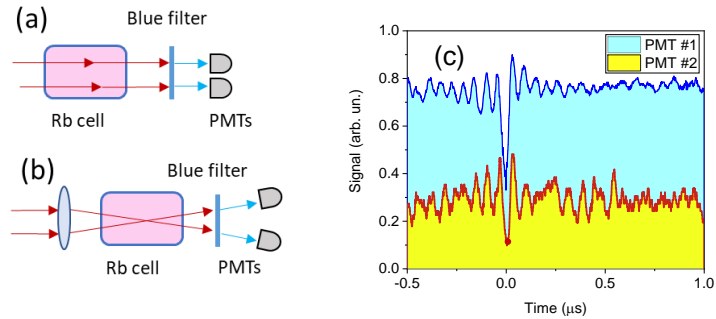

Fig. 2. (a, b) Optical schemes for studying intensity correlation of the directional blue emission at 420 nm in Rb vapors excited by two bi-chromatic laser beams. (c) Averaged temporal profiles of directional emission at 420 nm generated with the crossed pumping beams.

The observed weak anticorrelation might occur due to a gain depletion on the population-inverted transition by transversely-emitted spontaneous photons from the neighboring excitation region, while intensity correlation can be linked to synchronization by population-inverted atoms in the partially overlapping excitation regions.

## 3. Conclusion

Cooperative effects (CE) leading to superfluorescence or superradiance, were thoroughly studied in various systems with pulsed excitation [4,5]. We show that despite cw laser pumping, CE play an important role in determining temporal properties of the polychromatic frequency up- and down-converted emission. Short-pulse excitation is not a necessary condition for observing directional cooperative radiation. This reveals a new important aspect of collective emission.

A high level of intensity correlation is demonstrated not only for the 2338 nm and 330 nm fields coupled by FWM, but also between the cascade radiation at 2338 and 2207 nm. Correlation controlled by the geometry of the applied fields opens new possibilities for generating entangled fields from different spectral regions.